# A Study on the Impact of Gender, Employment Status and Academic Discipline on Cyber Hygiene: A Case Study of University of Nigeria, Nsukka

Celestine Ugwu[1*], Modesta Ezema[1], Uchenna Ome[1], Lizzy Ofusori[2], Comfort Olebera[3], and Elochukwu Ukwandu[4]

[1] Department of Computer Science, University of Nigeria, Nsukka
[2] School of Management, Information Technology & Governance, University of KwaZulu-Natal, Westvillle Campus, Durban, South Africa
[3] Dept. of Computer Science, Faculty of Physical Science, Imo State University, Owerri, Nigeria
[4] Dept. of Applied and Computing Engineering, Cardiff School of Technologies, Cardiff Metropolitan University, United Kingdom

(celestine.ugwu, modesta.ezema, uchenna.ome)@unn.edu.ng, lizzyofusori@yahoo.co.uk, Chy_prime@yahoo.com, eaukwandu@cardiffmet.ac.uk

Correspondent Author: Elochukwu Ukwandu, eaukwandu@cardiffmet.ac.uk

**Abstract.** The COVID-19 pandemic has helped amplify the importance of Cyber Hygiene. As the reliance on the Internet and IT services increased during the pandemic. This in turn has introduced a new wave of criminal activities such as cybercrimes. With the emergent of COVID-19 which lead to increase in cyber-attacks incidents, the pattern and sophistication, there is an urgent need to carry out an exploratory study to find out users' level of cyber-hygiene knowledge and culture based on gender, employment status and academic discipline. Above this, with many organisations providing for dual mode work pattern or remote and in-person as the pandemic subsides, this study remains very relevant and hence the aim to investigate the cyber-hygiene knowledge and compliance of university students and employees of the University of Nigeria, Nsukka (UNN). In addition, it attempts to verify the relationship between demographics such as gender, employment status and academic discipline on cyber hygiene culture among students and employees. The sample population is made of employees and students of UNN, where the employees are either academic staff or non-academic staff. The sample size consisted of three hundred and sixteen (316) participants, one hundred and eight-seven (187) of whom were females and one hundred and twenty-nine (129) were males. The results offer some useful insight on cyber hygiene practices at the university

## Introduction

Globally, many employees have shifted to work from home due to the COVID-19 pandemic [1]. The pandemic has disrupted business activities, work environments, and academic calendars, necessitating the use of Internet connectivity to work remotely.

Many organizations, including the educational sector, now rely on the Internet to perform basic and complex tasks [2]. Also, various academic institutions are currently making efforts to utilize Information and Communication Technology (ICT) tools and computer networks to support remote learning and online teaching in response to this pandemic [3,4]. However, in as much as the COVID-19 pandemic has necessitated the use of the Internet, ICT, and computer networks to work remotely, it has also introduced a new wave of criminal activity called cybercrime [5]. Cybercrime can be described as any illegal activity in which computers or computer networks are used as a tool to perform criminal activities [6]. Authors in [7] define it as those crimes committed through a computer or an electronic device, mainly through the Internet. According to [8], the Internet has become a vulnerable place where individuals, organizations, and agencies are constantly put at risk due to attacks by cybercriminals. The exploration rate of this internet vulnerability by online fraudsters seems to have increased as a result of COVID-19. In Nigeria, for instance, cybercrime has become one of the main avenues for pilfering money and business espionage [9].

Cybercrime activities have become worrisome with the emergence of the Internet particularly during the COVID-19 crisis [10]. Nonetheless, [4] suggests improving cyber-hygiene culture of the internet users as one of the various ways to curb cybercrime. According to [4], it is essential to investigate the effect of gender, employment status and area of discipline that might have impact on cyber hygiene and also take an assessment of cyber-hygiene culture among students and staff. While there have been studies on the influence of age and level of education among students and staff on cyber-hygiene, there is limited study on the impact of gender, employment status and academic discipline on cyber-hygiene [4]. This represents a gap in the literature, and it gives an opportunity for this study to address. Thus, this study aims to answer this research question on how gender, employment status and academic discipline impact cyber hygiene among students and employees of higher education institute, such as the University of Nigeria, Nsukka (UNN).

The study in addition, took an assessment of users' level of knowledge and habits towards cyber-hygiene at UNN. The contribution will add knowledge to literature, and the findings of the study may be compared to the findings in other related literatures.
The remainder of this paper is organized as follows: Section 2 is a review of the literature, while Section 3 discusses our research methods and data collection approaches. Section 4 shows the result of the data analysis, while Section 5 has the discussion of the results obtained and Section 6 is the concluding part of the work.

**Literature Review**
Cyber hygiene can be defined astaking precautions and implementing protocols that improve cybersecurity and maintain system health, thereby reducing the risks of falling victim to cyber-attacks [11]. Authors in [4] describe it as the precaution computer users take in keeping sensitive data safe, organized, and secure from theft and cyber-attacks.

As cyber-attacks increase rapidly, they are considered a major cyber security threat weakening the cybersecurity chain [12]. Cyber attacks challenge all network security regardless of the strength of their cryptography methods, anti-virus software programs, firewalls, and intrusion detection systems [12, 13]. Common cyber-attacks include malware attacks (worms, viruses, Trojan horse, Rootkit, ransomware attacks), distributed denial of service (DDOS), and phishing [12]. According to the study conducted by Omodunbi *et al.* (2016), 88% of the students are victims of phishing, while 65% are victims of Bank Verification Number scams. With this rapid increase of cyber-attacks, [9] affirm that cybercrime cannot be completely and easily wiped out but can be reduced.

Meanwhile, over the years, there have been enormous resources and funds allocated to cybersecurity; nevertheless, data breaches have also been on the increase [14,11]. This implies that the resources allocated to cybersecurity have not adequately curbed or prevented cyber-attacks. Several scholars have attributed the increase in data breaches/leakage to human error [14, 15, 11]. For example, a study conducted by [16] shows that 72% of people do not use firewall protection because they were not trained on the topic. Likewise, a recent study was carried out to understand the types of people who are more likely to engage in the risky behaviour of sharing passwords using the variables age, perseverance, and self-monitoring to understand [17]. The study's outcome reveals that the variables (age, perseverance, and self-monitoring) have a significant effect on the practice of sharing passwords [17]. However, [4] emphasize the importance of maintaining cyber-hygiene for internet users as not everyone cares enough to ensure that they are not hacked making cyber-hygiene one of the most overlooked necessities in modern society. It should be noted that a lot more is done online now than in the past hence ensuring that computer users maintain cyber-hygiene is key to ensuring that the computer systems and information are protected.

**Materials and Methods**
**Survey Participants**

Participants in our survey comprised employees and students of University of Nigeria, Nsukka, and the employees are either academic staff or non-academic staff. The sample consisted of three hundred and sixteen (316) participants, one hundred and eight-seven (187) of them were females and one hundred and twenty-nine (129) were males. In the aspect of their employment status, one hundred and sixty-nine (169) were students, one hundred (100) were academic staff, and forty-seven (47) were the non-academic staff. The distribution of the participants according to their academic discipline was as follows- two hundred and sixty (260) were science-based, fifty (50) were non-science based, and six (6) indicated neither science nor non-science as their discipline. The rationale behind the use of University of Nigeria for the study include for convenience, accessibility of ethical approval and availability of Wi-Fi network and also as academic at University of Nigeria, we wanted to understand the importance of good cyber-hygiene so as to advise the institution correctly. Ethical approval was given for this study by the university authority.

**Research Goal and Procedure**

This research aims at investigating the cyber-hygiene knowledge and compliance of university students and employees. In addition, it attempts to verify the relationship between demographics such as gender, employment status and academic discipline on cyber hygiene culture among students and employees. University of Nigeria Nsukka was used as a case study.

Prior to collecting data from our prospective participants, ethical approval for the study was sought from the case study institution, University of Nigeria, Nsukka, and it was obtained. Data collection was done through a structured questionnaire. The questionnaire administered to the participants with the aid of Google form. Google form was utilized because this research was carried out during COVID-19 pandemic that requires people comply with the COVID-19 regulations. A non-probability sampling technique, the convenience strategies was used in choosing this study's prospective participants. A total of 316 responses were received and analyzed using a statistical analysis tool that examines the relationship of variables that are considered in this research work.

**Research Hypothesis**

The following null hypotheses were the made for the study:

**H1**: Gender of internet users does not have significant effect on the cyber-hygiene

**H2**: Employment status of internet users does not have significant effect on cyber-hygiene

**H3:** Academic discipline of internet users does not have significant effect on cyber-hygiene

**Data Analysis**

The questionnaire consisted of three sections: demographic, knowledge of threat and concept, and cyber-hygiene attitude. The first section gathered information on the participants demographic factors, while the second and the third section assessed the participants knowledge of threats/concepts and cyber-hygiene attitudes. The questionnaire contained questions drawn from the literature on current cyber-hygiene practices such as storage and virus, social network, authentication, and social engineering. The second section consisted of 12 multiple questions while the third section comprised of 13 questions, some of the questions using likert-type answer formats with responses ranging from "every time" to "never". The likert-type answer formats were further reduced to binary-type response by considering either "every time or often" as correct response and "rarely and never" as wrong answer or "rarely and never" as correct answer and "every time and often" as wrong answer. This can be justified by facts that some of the questions accept either "every time" or "often" as correct answer while some questions accept either "rarely" or "never" as correct answer in line with good cyber-hygiene practices.

The data analysis was performed using analyzing software known as Statistical Package for Social Sciences (SPSS version 20). The analysis results are as shown in the next section of the paper.

**Result**

The survey sought information on demographic data, knowledge of threats and concepts of cyber-hygiene and cyber-hygiene culture.

Initially, percentage analysis, tables, and pie-charts were used to analyze and interpret data. Based on the participants' responses to the questionnaire, descriptive statistics were used to present the findings in this study. Later, a chi-square test was used to ascertain the association between respondent's demographic factors (gender, employment status and academic discipline) and cyber-hygiene culture. Chi-square is a good and simple statistical tool for verifying association among variables.

**Descriptive Statistics of Demographics**

Using descriptive statistics for the analysis, some of the data associated with the first section of the questionnaire (demographic) were presented as follows. Table 1 shows the distribution of participants according to their gender, employment status, and academic discipline.

Table 1: Socio-Demographic Data

| Variable | Frequency | Percentage (%) |
|---|---|---|
| **Gender** | | |
| Female | 187 | 59.2 |
| Male | 129 | 40.8 |
| **Employment Status** | | |
| Student | 169 | 53.5 |
| Academic staff | 100 | 31.6 |
| Non-Academic staff | 47 | 14.9 |
| **Academic Discipline** | | |
| Science | 260 | 82.3 |
| Non-Science | 50 | 15.8 |
| Neither | 6 | 1.9 |

**Descriptive Statistics of Cyber-hygiene Knowledge Concept and Threats**

This section presents data analysis obtained from eight out of twelve questions in the second part of the questionnaire. Based on the participants' responses to the questionnaire, some parts of the data associated with the second section (knowledge of threats and concepts of cyber-hygiene) were analysed, and the following Pie charts were formed.

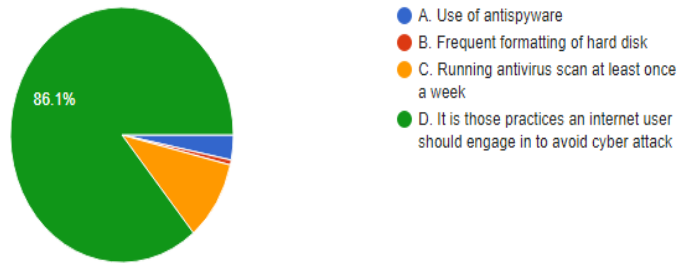

**Figure 1: Understanding the Meaning of Cyber-Hygiene**

Figure 1 shows the percentage of responses given to question number 1 of section 2; "What do you understand by cyber-hygiene". The pie chart shows that 86.1% of the participants have knowledge of what cyber-hygiene is all about.

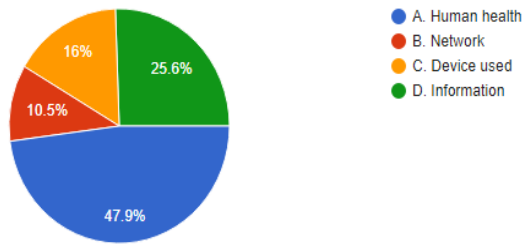

**Figure 2: Knowledge on the effect of poor cyber-hygiene**

Figure 2 presents the percentage of responses given to question number 2 of section 2; "The following could be affected as a result of poor cyber-hygiene except." The result shows that 47.9% know the likely effect of poor cyber-hygiene.

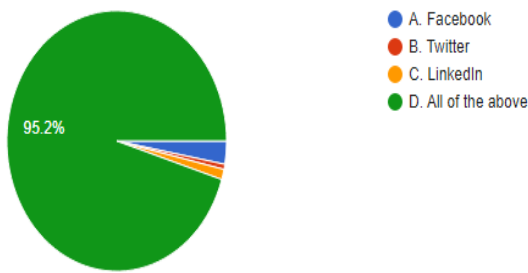

**Figure 3: Knowledge of social networking**

Figure 3 presents the percentage of responses given to question number 3 of section 2; "Which of the following is an example of social networking." From the result, 95.2% of the respondents are aware of the existing social networking.

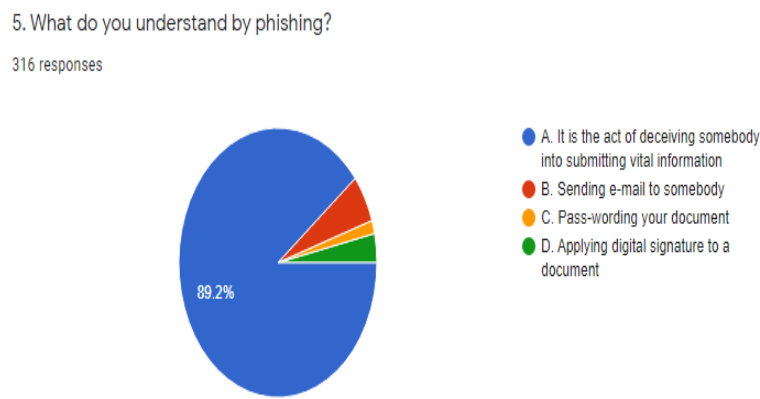

**Figure 4: Knowledge of phishing**

Figure 4 presents the percentage of responses given to question number 5 of section 2; "What do you understand by phishing." This result shows that 89.2% of the participants have knowledge of what phishing is.

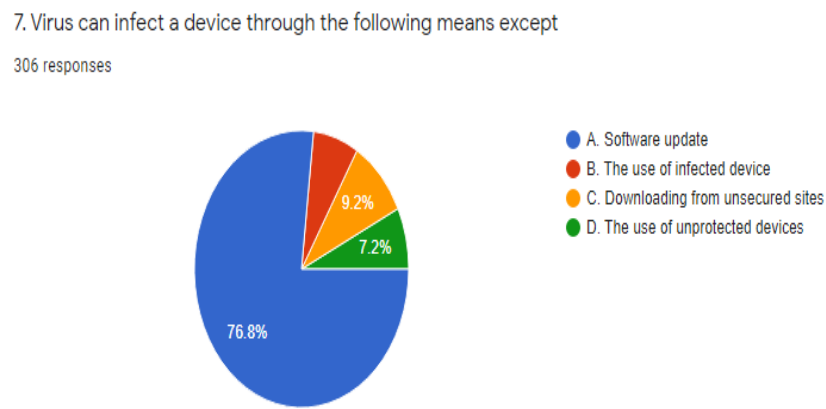

**Figure 5: Knowledge of means transmitting virus**

Figure 5 presents the percentage of responses given to question number 7 of section 2; "Virus can infect a device through the following means except." Figure 1 shows that 76.8% of the participants are aware of the possible means of transmitting virus.

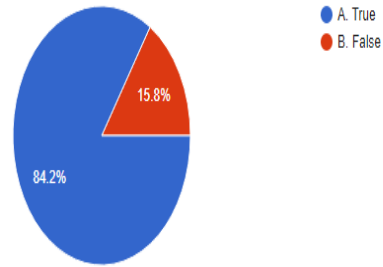

**Figure 6: Knowledge of antivirus**

Figure 6 presents the percentage of responses given to question number 8 of section 2; "Firewall is also a kind of antivirus." Figure 6 shows that 84.2% of the respondents know firewall as a kind of antivirus.

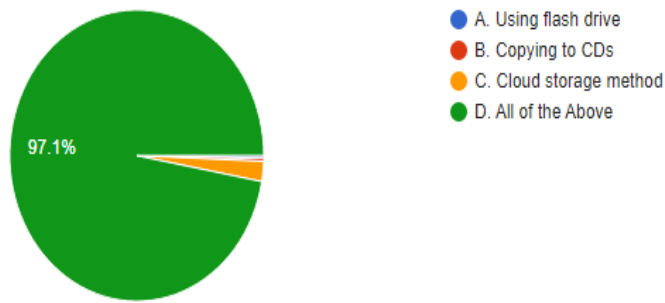

**Figure 7: Knowledge of storage**

Figure 7 presents the percentage of responses given to question number 10 of section 2; "Ways of backing up files include." Figure 7 shows that 97.1% of the respondents are aware of the possible techniques used for file backup.

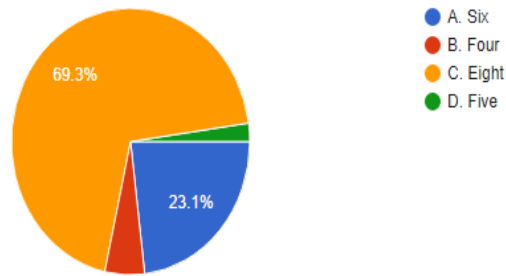

**Figure 8: Knowledge of Password**

Figure 8 presents the percentage of responses given to question number 11 of section 2; "The maximum number of characters that a strong password should contain is." Figure8 shows that 69.3% aware of the number of characters needed to form strong password.

**Descriptive Statistics of Cyber-hygiene Culture**

The descriptive statistics of eight out of thirteen questions in section 3 are presented as follows.

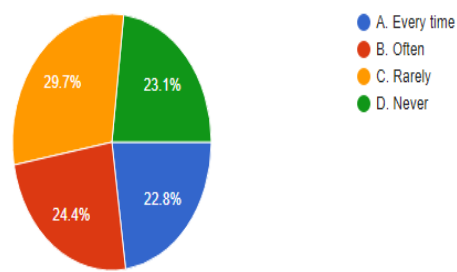

**Figure 9: Usage of Antispyware**

Figure 9 presents the percentage of responses given to question number 2 of section 3; "When do you use antispyware software." Figure 9 shows that a total of 52.8% selected "every time" and "often" as their culture in using antispyware.

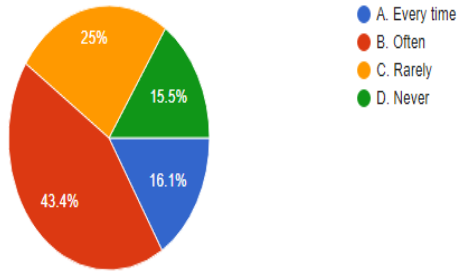

Figure 10: Usage of Password

Figure 10 presents the percentage of responses given to question number 5 of section 3; "How often do you use single password for multiple accounts." The result in Figure10 shows that 59.5% selected "every time" and "often" as their culture in using single password for multiple accounts.

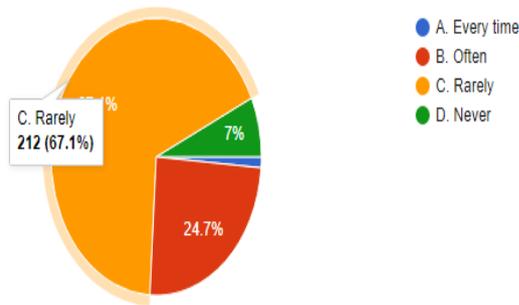

**Figure 11: Change of Password**

Figure 11 presents the percentage of responses given to question number 6 of section 3; "How often do you change your password." Figure11 indicates that 25.9% selected "every time" and "often" as their culture of changing password.

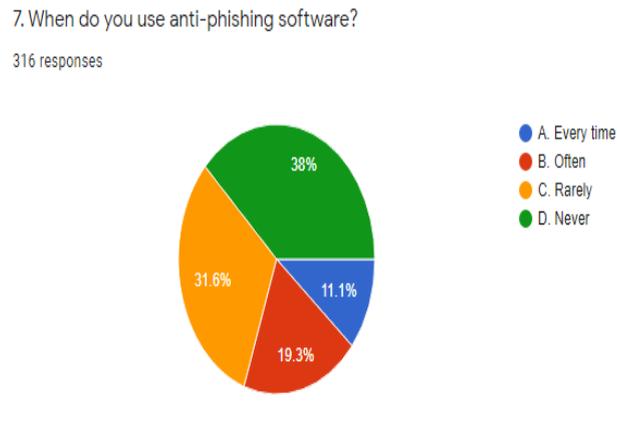

**Figure 12: Usage of Anti-Phishing Software**

Figure 12 presents the percentage of responses given to question number 7 of section 3; "When do you use anti-phishing software." It was shown in Figure12 that 30.4% selected "every time" and "often" as their culture of using ant-phishing software.

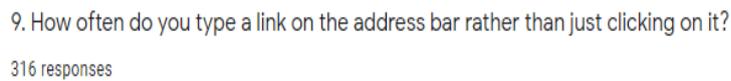

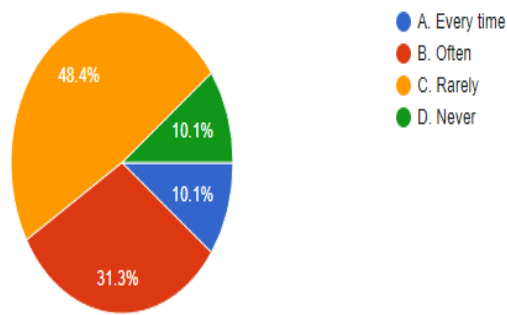

**Figure 13: Avoidance of Phishing Threat**

Figure 13 presents the percentage of responses given to question number 9 of section 3; "How often do you type a link on the address bar rather than just clicking on it." Figure 13 shows that 41.4% of the participants selected "every time" and "often" as their culture of typing address instead of clicking on the link.

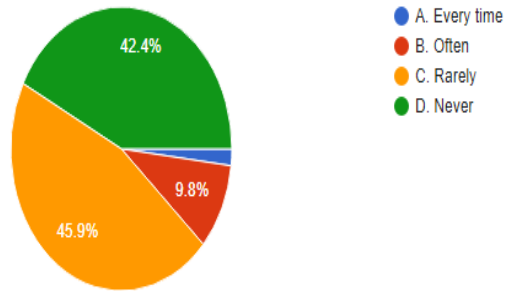

**Figure 14: Avoidance of Social Engineering Threat**

Figure 14 presents the percentage of responses given to question number 10 of section 3; "How often do you respond to emails or links that are asking for sensitive information." The result in Figure 14 shows that 11.7% selected "every time" and "often" as their culture of responding to links and emails requesting for sensitive information.

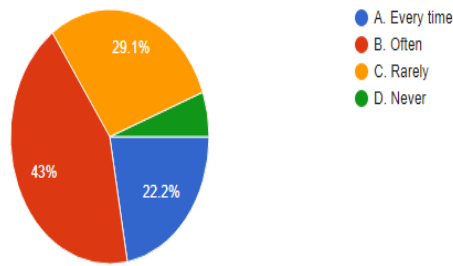

**Figure 15: Hiding Real Information on Social Media**

Figure 15 presents the percentage of responses given to question number 11 of section 3; "How often do you use any of your real name, email address and date of birth on social media." Figure 15 shows that 65.2% of the respondents selected "every time" and "often" as their culture of supplying real information on social media.

**Demographics and Cyber-hygiene Culture**

The variables representing questions in cyber hygiene culture domain were measured using interval scale; hence the variables were added to get Total Cyber Hygiene Culture (TCHC). The average was calculated by dividing the total by the number of questions in the domain. The mean was used ad cut-off for categorized Average Total Cyber Hygiene Culture (CatATCHC). Respondents who scored an average of the mean and above were rated as having high cyber hygiene culture, and given nominal value of 1, while those with average below the mean were rated as having low cyber hygiene culture and assigned nominal value of 0.

**Gender and Cyber-hygiene Culture**

Analysis of gender-based responses with respect to cyber hygiene culture showed that a total of 187 (89.2%) male respondents participated in the study, out of which 105 (66%) were found to have poor cyber hygiene culture, while the remaining 82 (52.2%) were found to have good cyber hygiene culture. The female category had 129 (40.8%) participants, out of which 54 (34.0%) were rated as having low cyber hygiene culture while the remaining 75 (47.8%) were rated as having good cyber hygiene culture. The crosstabulation result is presented in table 2 below:

**Table 2: Gender * Categorized average total cyber hygiene culture Crosstabulation**

| | | | Categorized average total cyber hygiene culture | | Total |
|---|---|---|---|---|---|
| | | | Poor | good | |
| Gender | Male | Count | 105 | 82 | 187 |
| | | %within categorized average total cyber hygiene culture | 66.0% | 52.2% | 59.2% |
| | Female | Count | 54 | 75 | 129 |
| | | %within categorized average total cyber hygiene culture | 34.0% | 47.8% | 40.8% |
| Total | | Count | 159 | 157 | 316 |
| | | %within categorized average total cyber hygiene culture | 100.0% | 100.0% | 100.0% |

**Table 3: Chi-Square Tests for Gender and Categorized Average Total Cyber Hygiene Culture**

| | Value | Df | Asymp. Sig. (2-sided) | Exact Sig. (2-sided) | Exact Sig. (1-sided) |
|---|---|---|---|---|---|
| Pearson Chi-Square | 6.235 | 1 | .013 | | |
| Continuity Correction | 5.677 | 1 | .017 | | |
| Likelihood Ratio | 6.257 | 1 | .012 | | |
| Fisher's Exact Test | | | | .016 | .009 |
| Linear-by-Linear Association | 6.215 | 1 | .013 | | |
| N of Valid Cases | 316 | | | | |

Since the p-value of 0.13 is greater than the statistically accepted significant value of <=0.05, we conclude that there is no significant relationship between the gender of internet users and their cyber hygiene culture.

**Employment Status and Cyber-hygiene Culture**

Analysis of the relationship between the employment status of internet users and their cyber hygiene culture showed that 169 (53.5%) respondents in the student category participated in the study. Out of which, 90 (56.6%) were found to have poor cyber hygiene culture, while 79 respondents (50.3%) were found to have good cyber hygiene culture. Out of 100 (31.6%) respondents in the academic staff category, 41 (25.8%) were found to have poor cyber hygiene culture, while the remaining 59 (37.6%) were found to have good cyber hygiene culture. In the Non-academic staff category, out of 47(14.9%) who participated in the study, 28 (17.6%) were found to have poor cyber hygiene culture, while 19 (12.1%) were found to have high cyber hygiene score. The result of the analysis is shown in table 4 below:

**Table 4: Status * Categorized average total cyber hygiene culture Crosstabulation**

| | | | Categorized average total cyber hygiene culture | | |
|---|---|---|---|---|---|
| | | | poor | good | Total |
| Status | Student | Count | 90 | 79 | 169 |
| | | % within categorized average total cyber hygiene culture | 56.6% | 50.3% | 53.5% |
| | Academic staff | Count | 41 | 59 | 100 |
| | | % within categorized average total cyber hygiene culture | 25.8% | 37.6% | 31.6% |
| | Non-academic staff | Count | 28 | 19 | 47 |
| | | % within categorized average total cyber hygiene culture | 17.6% | 12.1% | 14.9% |
| Total | | Count | 159 | 157 | 316 |
| | | % within categorized average total cyber hygiene culture | 100.0% | 100.0% | 100.0% |

To find a significant relationship between Employment Status of internet users and cyber hygiene culture, categorical variable of employment status and the categorical variable on cyber hygiene culture were subjected to the Chi-Square test. The result is presented in the Table 5:

**Table 5: Chi-Square Tests for Employment status and cyber hygiene culture relationship**

|  | Value | Df | Asymp. Sig. (2-sided) |
|---|---|---|---|
| Pearson Chi-Square | 5.667 | 2 | .059 |
| Likelihood Ratio | 5.696 | 2 | .058 |
| Linear-by-Linear Association | .009 | 1 | .925 |
| N of Valid Cases | 316 |  |  |

Since the p-value of 0.059 is slightly greater than the statistically accepted significant value of <=0.05, we conclude that there is no significant relationship between the employment status of internet users and their cyber hygiene culture.

**Academic Discipline and Cyber-hygiene**

Analysis of categorized variables of education discipline and cyber hygiene culture showed that 260 respondents in the Science category (82.3%) participated in the study. It was found the 128 (80.5%) have poor cyber hygiene culture, while 132 (84.1%) have good cyber hygiene culture. In the Non-Science category, 50 (15.8%) participated in the study where 27 (17.0%) were found to have poor cyber hygiene culture, while the remaining 23 (14.6%) were found to have good cyber hygiene culture. It was also found that 6(1.9%) in neither the Science category nor the non-Science category participated in the study and 4(2.5%) of respondents have poor cyber hygiene culture while the remaining 2 (1.3%) have good cyber hygiene culture. The crosstabulation result is presented in table 6 below:

**Table 6: Academic Discipline * Categorized average total cyber hygiene culture Crosstabulation**

|  |  |  | Categorized average total cyber hygiene culture | | Total |
|---|---|---|---|---|---|
|  |  |  | poor | good |  |
| Academic Discipline | Science | Count | 128 | 132 | 260 |
|  |  | % within categorized average total cyber hygiene culture | 80.5% | 84.1% | 82.3% |
|  | Non-Science | Count | 27 | 23 | 50 |
|  |  | % within categorized average total cyber hygiene culture | 17.0% | 14.6% | 15.8% |
|  | Neither | Count | 4 | 2 | 6 |
|  |  | % within categorized average total cyber hygiene culture | 2.5% | 1.3% | 1.9% |
| Total |  | Count | 159 | 157 | 316 |
|  |  | % within categorized average total cyber hygiene culture | 100.0% | 100.0% | 100.0% |

To find if there exists a significant relationship between Academic discipline of internet users and cyber hygiene culture, categorical variables of academic discipline and categorical variables on cyber hygiene culture were subjected to the Chi-Square test. The result is presented in Table 7 below:

**Table 7: Chi-Square Tests of Academic discipline and Cyber Hygiene Culture Relationship**

|  | Value | Df | Asymp. Sig. (2-sided) |
|---|---|---|---|
| Pearson Chi-Square | 1.036 | 2 | .596 |
| Likelihood Ratio | 1.049 | 2 | .592 |
| Linear-by-Linear Association | .933 | 1 | .334 |
| N of Valid Cases | 316 |  |  |

Since the p-value of 0.596 is greater than the statistically accepted significant value of <=0.05, we conclude that there is no significant relationship between the Academic discipline of internet users and their cyber hygiene culture.

**Discussion**

The study investigated the impact of gender, employment status and academic discipline on cyber-hygiene culture. The descriptive statistics of the result of assessment of users' knowledge and culture about cyber-hygiene were presented. It was found that internet users' level of knowledge about cyber-hygiene is higher when compared to their culture or habit towards cyber-hygiene. Apart from the likely effect of poor cyber-hygiene where the assessment finding is 47.9%, other findings on their level of knowledge range from 69.3% to 97.1% as shown in Figure 1 to Figure 8.

According to our findings on our respondents' cyber-hygiene culture, it was found that 52.8% (see Figure 9) of the participants use antispyware software, which does not follow the finding of 97% from [16] but allies with the finding of 47% - 78% from [18]. Findings also showed that 59.5% of internet users have the habit of using a single password for multiple accounts and only 25.9% have the habit of changing their password regularly; thus, 74.1% do not change it often (see Figures 10 and 11). It was found that 30.4% of the participants use anti-phishing software, 41.4% regularly type addresses rather than clicking, and only 11.7% respond to emails and links requesting for sensitive information (see Figures 12, 13, and 14). The 11.7% disagree with the finding of 88% of students being victims of phishing as found by [9]. The reason for the discrepancy may be because of the level of awareness made on phishing threats within the few years of the interval between the two studies. It also found from our study that 65.2% of users supply their real information on social media (see Figure 15).

In an attempt to investigate the impact certain attributes of internet users, we made discoveries that are insightful. It is commonly believed that students usually portray poor cyber-hygiene knowledge and culture because of lack of experience and training when compared to the working class. According to [19], there is need to provide awareness and education to students who are the potential targets for cyber exploitation. The findings from [19] suggest that majority of students may lack an understanding of the importance of cybersecurity. But surprisingly, result showed that employment status (students and working-class) has no statistical significance to cyber-hygiene culture. This finding is likely to be true since most of the employees are neither engaged in training nor awareness programmes in the aspect of cyber-hygiene as expected.

It has been revealed in previous studies that gender is an attribute that could have effect on people's cybersecurity behaviour [21]. It is presumed that males use the internet more often than their female counterparts and as such, they are expected to exhibit better culture towards cyber-hygiene. However, [20] found that the difference of gender does not show any difference in the frequency of internet usage for downloading, social networking, chatting, and purchasing. Also, contrary to the belief, the finding from our study indicated that gender has no statistical significance on cyber-hygiene culture. This study's findings agree with findings from [18] that males did not differ on cyber-hygiene behavior from females, however, this study's findings differ with the findings from [21], that disparity exists between the female and male students regarding their security self-efficacy.

We also examined whether being science inclined or not has an impact on the cyber-hygiene culture. The study's finding showed that academic discipline has no statistical significance on the cyber-hygiene even when it is believed that those in area of sciences are more friendly with internet and as such should use it more securely.

**Conclusion and Future Work**

This study took a detailed assessment of cyber-hygiene knowledge and culture, and how certain internet users attribute impact on their cyber-hygiene culture. Our findings agree and disagree with some previous research in this domain of study. The study also has findings that provide further insight on the poor knowledge and culture of internet users towards cyber-hygiene. Thus, we suggest the providing elaborate awareness and training for both students and staff of universities of underdeveloped countries to curb the menace of cyber threat. The global security organization, Cloud Security Alliance (CSA), also believe that most of the cybersecurity incidence occur due to lack of awareness and cited social engineering as the most common reason [19].

In line with the recommendation for an elaborate awareness and training as suggested in this study and other related studies, the future direction for study should be on the design and implementation of an acceptable template for awareness creation and training of students and staff of universities. According to [22], most of the research conducted globally on internet use has findings that internet usage is most prevalent among younger and more educated people. We also suggest the application of machine learning to make prediction on the cyber-hygiene knowledge and culture of students and staff using their responses as another area for future work.